\documentclass[letter,12pt]{article}
\usepackage{graphicx}
\usepackage{hyperref}
\usepackage{upgreek}
\usepackage{multicol,multirow}
\usepackage{amsmath,amssymb,amsfonts}
\usepackage{mathrsfs}
\usepackage{amsthm}
\usepackage[figuresright]{rotating}
\usepackage{appendix}
\usepackage{ifpdf}
\usepackage[T1]{fontenc}
\usepackage{newtxtext}
\usepackage{newtxmath}
\usepackage{textcomp}
\usepackage{authblk}
\usepackage[left=1in,right=1in]{geometry}
\usepackage{xcolor}
\usepackage{subcaption}
\usepackage{apacite}    
\usepackage[round]{natbib}

\newcommand{\St}{\mathrm{St}}
\renewcommand{\Re}{\mathrm{Re}}
\newcommand{\sgn}{\operatorname{sgn}}

\title{Beaching model for buoyant marine debris in bore-driven swash}
\author[$^1$]{Benjamin Davidson}
\author[$^1$]{Jamie Brenner}
\author[$^1$]{Nimish Pujara}
\affil[$^1$]{Department of Civil and Environmental Engineering, University of Wisconsin--Madison, Madison WI 53706, USA}

\begin{document}
\date{}
\maketitle

\begin{abstract}
Marine debris pollution is a growing problem impacting aquatic ecosystems, coastal recreation, and human society. Beaches are known to be a sink for debris, and beaching needs to be accounted for in marine debris mass balances.  The process of buoyant debris beaching is not sufficiently well understood in order to include this process yet.  We develop a simplified model for buoyant marine debris transport in bore-driven swash (where the water wets the beach with each incoming wave). We validate the model with laboratory experiments and use the combined results from the model and experiments to understand the parameters that are important for dictating particle beaching. The most relevant parameters are the particle inertia and the timing and velocity with which debris particles enter the swash zone.
\end{abstract}

\maketitle

\section{Introduction}\label{sec-introduction}

It is no surprise that plastics are the largest component of global marine litter given that plastic waste generation is still increasing, and is likely to continue to increase with the current trends of plastic consumption \citep{geyer_production_2017}.  In one year it is estimated that 11\% of plastic waste can make its way into marine environments \citep{borrelle_predicted_2020}. Some unknown fraction of that plastic debris collects on shorelines around the world \citep{rosevelt_marine_2013}, with some trends pointing to more marine debris accumulation nearer to population centers \citep{galgani_global_2015}.  In studies of marine debris sampled on beaches, plastics often dominate regardless of the debris size or beach location \citep{haseler_marine_2020, eriksson_daily_2013, compa_spatial_2022, rios_spatio-temporal_2018}.  Across an aggregate of almost 1,500 publications concerning aquatic debris pollution globally, macro-debris on beaches has been shown to be mostly plastic while beached micro-debris was found to be almost entirely composed of plastics \citep{litterbase}.  Thus, beaches are clearly collection sites for marine debris, but it is unknown at what rate and efficiency this collection occurs.  

Due to the vast size and dynamic nature of oceanic circulation, numerical models prove to be an effective tool for understanding plastic debris fate and transport throughout the marine environment.  Effective models must account for debris sources and sinks, specifically including shoreline interaction and beaching.  Current models of plastic debris have focused on medium-to-large-scale marine models with poorly defined beaching dynamics, some ignoring debris beaching altogether \citep{van_sebille_origin_2012}.  For example, \citet{neumann_marine_2014} did not consider beaching but rather considered the shoreline to be a reflective boundary in the model. \citet{jalon-rojas_3d_2019} and others, include a parameter for particles to wash off the beach and be re-suspended without including much detail on debris deposition on the beach.  When concerning beach deposition, a common beaching parameter across multiple models was to consider a particle deposited on a beach if the particle was in a computational cell along the coast \citep{collins_modelling_2019,lebreton_numerical_2012,atwood_coastal_2019, lebreton_global_2019}.  This is sometimes accompanied by a requirement for the particle to be stagnant for a duration of time. This is problematic since the resolution of these models is usually on the order of kilometers, which is orders of magnitude larger than the characteristic size of the particles themselves and the surf and swash zone dynamics that determine particle beaching. Clearly, coarse-resolution models require parameterization of beaching, but these can be made more accurate with a better understanding of smaller-scale dynamics. We tackle this need to understand debris beaching, which can be used to better represent this phenomenon in mass budget models of plastic debris transport throughout the environment.

The question of microplastic beaching has been studied previously with modeling showing that some of the important factors to consider are the sources and quantity of plastic debris and the re-suspension of plastics from beaches \citep{critchell_modelling_2016}.  In order to understand plastic debris re-suspension, a better understanding of initial particle beaching at the scale of the swash processes is necessary.  \citet{hinata_numerical_2020} develop a probabilistic plastic beaching model based on the fluxes of deposited and suspended plastics with the particle residence time on the beach.  However, this model focuses on the drift of particles onto the beach from the nearshore without much consideration for turbulent swash processes.  This has been experimentally investigated in the context of wind and wave forcing by \citet{forsberg_behaviour_2020} where they saw buoyant particles mainly deposited on the beach after reaching a steady state under regular waves.  In contrast, \citet{kerpen_wave-induced_2020} found buoyant plastics to be distributed nearly equally between the beach and water surface after continuous wave forcing in laboratory experiments. \citet{nunez_wave-induced_2023} experimentally investigated the fate of particles whether originating on or offshore, of many plastic types and shapes, and across various wave conditions.  \citet{larsen_experimental_2023} measured the time it took for microplastic particles to beach in laboratory experiments given various starting conditions and empirically modeled the particle transport.  We begin to understand buoyant plastic beaching through these models and experiments, yet the physical dynamics leading to the particle deposition on the beach are still unclear.

In this work, we consider the characteristics which influence the deposition of buoyant debris on a shallow beach in the context of bore-driven swash flow using a combination of mathematical modeling and laboratory experiments.  We account for the conditions of the particle as it approaches the shore and the particle inertia.  We find the velocity of the particle relative to the bore celerity and the lag of the particle behind the bore front to be important parameters in addition to the particle inertia.  We find beaching is more common for the particles with larger inertia and those that enter the swash zone closer to the bore front in time and speed.

 In $\S$\ref{sec-model}, we develop the particle motion model for buoyant disks in the swash zone, specifically considering particle beaching.  We explore the parameters influencing particle fate in regard to deposition on the beach.  This model is compared with accompanying laboratory experiments in $\S$\ref{sec-experiments}.  Finally, we explore how the forces on buoyant particles dictate beaching in the context of the experimental and model results, including the sensitivity of the model results to its input parameters in $\S$\ref{sec-discussion}. We offer brief concluding remarks in $\S$\ref{sec-conclusions}.

\section{Model}\label{sec-model}

    \subsection{Equation of particle motion for inertial floating disks in the swash zone}\label{sec-particle-equation}
        We consider the forces acting on an inertial, disk-shaped particle that is floating during a single wave swash event on an inclined slope using a framework from \citet{maxey_equation_1983}. Expanding the work from \citet{maxey_equation_1983} to include a free surface has been shown previously by \citet{rumer_ice_1979} with regular waves and \citet{beron-vera_building_2019} with surface currents.  The work by \citet{rumer_ice_1979} has been expanded by others \citep{shen_theoretical_2001, calvert_mechanism_2021, huang_analytical_2016} and considers the object motion to be driven by gravity; the particle slides down the slope of a surface wave in a reference frame that rotates with the wave motion.   We focus on the dynamics of the swash zone, the region from the maximum wave run-up to the minimum wave draw-down on a beach slope.  In bore-driven swash, the slope of the free surface is small \citep{pujara_swash_2015}, so we maintain a constant reference frame and consider the particle motion to be driven by the current of the swash up the beach, more similar to \citet{beron-vera_building_2019}.

\begin{figure}
            \centering
            \includegraphics[width=1.0\textwidth]{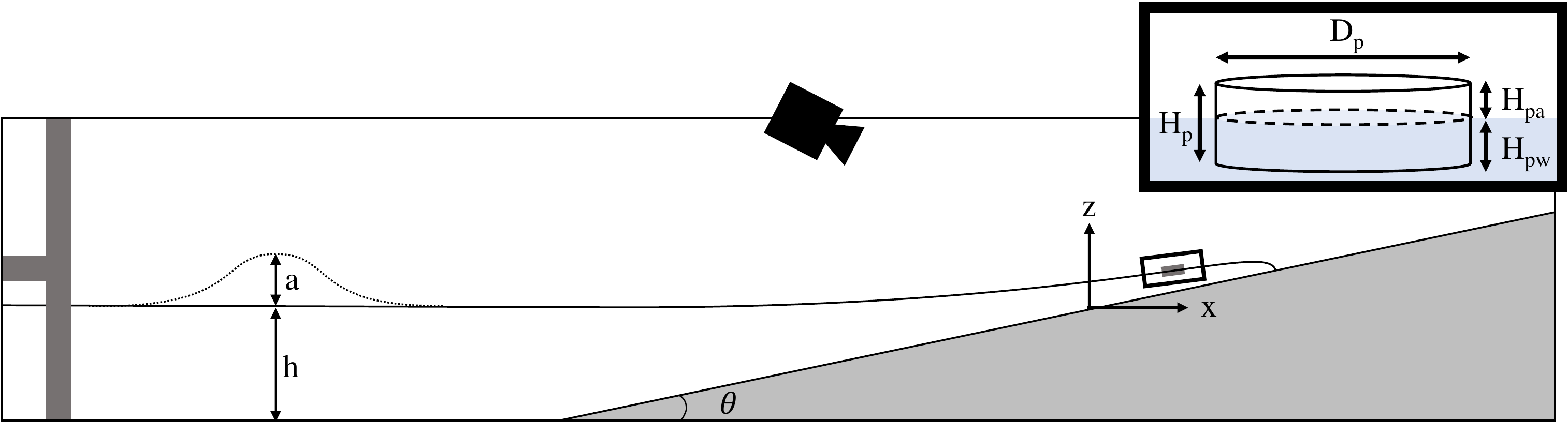}
            \caption{\label{Fig1} Definition sketch of the incident wave and resulting swash flow. The experimental set-up consists of a piston-type wave maker and a plane, impermeable beach with angle $\tan \theta = 1/10$, and an oblique overhead camera pointing at the beach. The still water depth is represented by $h$ with the water wave amplitude $a$ in the constant depth region. The solid line indicates the water profile during the wave run-up. The buoyant particle is a cylindrical disk with diameter $D_{p}$, height $H_{p}$, submerged height $H_{pw}$, and height above the water surface of $H_{pa}$.}
        \end{figure}

        Figure \ref{Fig1} shows the setup, undisturbed wave characteristics, and a particle floating in the swash zone during a wave event.  We define an $x-z$ coordinate system whose origin is at the undisturbed shoreline with $x$ pointing horizontally towards the shore and $z$ vertically. We consider the dynamics of a floating disk-shaped particle with diameter $D_{p}$, height $H_{p}$, submerged height $H_{pw}$, and height above the water surface $H_{pa}$. Considering only the forces on the particle from the water (and ignoring forces acting on the particle in the air), we formulate a particle equation of motion that considers the particle to be in vertical equilibrium and subject to accelerations in the horizontal direction due to different forces. 

        When the particle is in deep enough water to not touch the bottom, the vertical force balance is simply the weight of the particle opposed by buoyancy.   These forces are related by the specific gravity ($\gamma$) of the particle, which is the ratio of the particle density ($\rho_p$) to the fluid density ($\rho$).  $\gamma$ is also equal to the fraction of the volume of fluid displaced by the particle over the volume of the particle in water deep enough to not be touching the beach surface.  The disk shape simplifies the analysis significantly compared with the sphere considered previously \citep{beron-vera_building_2019} since the submerged particle depth is linearly proportional to the volume of water displaced.  We can simplify $\gamma$ to a ratio of the submerged particle height to the total particle height which we define as $\phi$,
        \begin{equation} \label{eq-gamma-phi}
            \gamma = \frac{\rho_p}{\rho} = \frac{\frac{\pi}{4}D_p^2H_{pw}}{\frac{\pi}{4}D_p^2H_{p}} = \frac{H_{pw}}{H_{p}} = \phi.
        \end{equation}


        Our model differs from those noted above and expands equation \eqref{eq-gamma-phi} to include a reaction force of the particle on the beach when considering of the forces in the $z$-direction.  For the partially submerged particle that could be touching the beach, we consider a ratio of the mass of fluid displaced to the mass of the particle, which simplifies to
        \begin{equation}\label{eq-mass-ratio}
            \frac{m_f}{m_p} = \frac{\rho}{\rho_p}\frac{\frac{\pi}{4}D_p^2 H_{pw}}{\frac{\pi}{4}D_p^2 H_p} = \frac{\rho}{\rho_p}\frac{H_{pw}}{H_p} = \frac{\phi}{\gamma}.
        \end{equation}

        For the particle motion up the beach, we consider the acceleration in the $x$-direction, which is different to the direction up the slope of the beach by a factor of $\cos \theta$.  Because the angle $\theta$ tends to be small, the difference between the horizontal and beach-parallel directions is negligible. Consider that $\cos\theta = 1- \tfrac{1}{2}\theta^2 + \mathcal{O}(\theta^3)$ and $\tan\theta = \theta + \mathcal{O}(\theta^3)$ for small $\theta$.  Using  $s=\tan(\theta)$ to define the slope, we see that $\cos\theta \approx 1$ with a leading order error of $\tfrac{1}{2}s^2$.  For a beach slope of $s=1/10$, this gives a leading order error from using the small angle approximation of $5 \times 10^{-3}$.

        The force on the particle up the beach slope is a combination of the forcing of the fluid, the forcing of the added mass from the fluid, the force of drag from the fluid acting on the particle, and frictional forces with the beach.  This is expressed dimensionally as:
        \begin{equation}\label{eq-particle-motion-dimensional}
            m_p\frac{dv'_p}{dt'} = m_f \frac{Du'}{Dt'} + C_m m_f \left( \frac{Du'}{Dt'} - \frac{dv'_p}{dt'} \right) -C_D \frac{1}{2}\rho |v'_p - u'|(v'_p-u')D_p H_{pw} + \left[\; F'_F\;\right],
        \end{equation}
        where $m_p$ is the mass of the particle, $m_f$ is the mass of fluid displaced by the particle, $v'_p$ is the velocity of the particle, $u'$ is the fluid velocity, $C_m$ is the coefficient of added mass, $C_D$ is the coefficient of drag, and $F'_F$ is the force of friction with the beach (which only acts when the water depth is sufficiently small that the particle makes contact with the beach).  Here we note that the $v'_p$, $u'$, $t'$, and $F'_F$ are dimensional variables indicated by the prime notation.

        We divide equation \eqref{eq-particle-motion-dimensional} by $m_p$ using equation \eqref{eq-mass-ratio}, and simplify to isolate the particle acceleration.  We let the acceleration due to friction be $a_F' = F'_F/m_p$ since this term is only active when the water depth is negligibly small.  We address the frictional acceleration fully in $\S$ \ref{sec-beaching-considerations}. We define a Reynolds number of the flow by the particle slip velocity and particle diameter as $\Re =|v_p' - u'|D_p/\nu$, resulting in
        \begin{equation}\label{eq-particle_2}
             \frac{dv'_p}{dt'}  = \left( \frac{\phi + C_m \phi}{\gamma + C_m \phi} \right) \frac{Du'}{Dt'}  - \left( \frac{1}{\gamma + C_m \phi} \right)\frac{2 C_D \phi \nu}{ \pi D_p^2 }\Re(v'_p-u') + \left[ \; a'_F \;\right].
        \end{equation}

\begin{figure}
            \centering
            \includegraphics[width=0.4\textwidth]{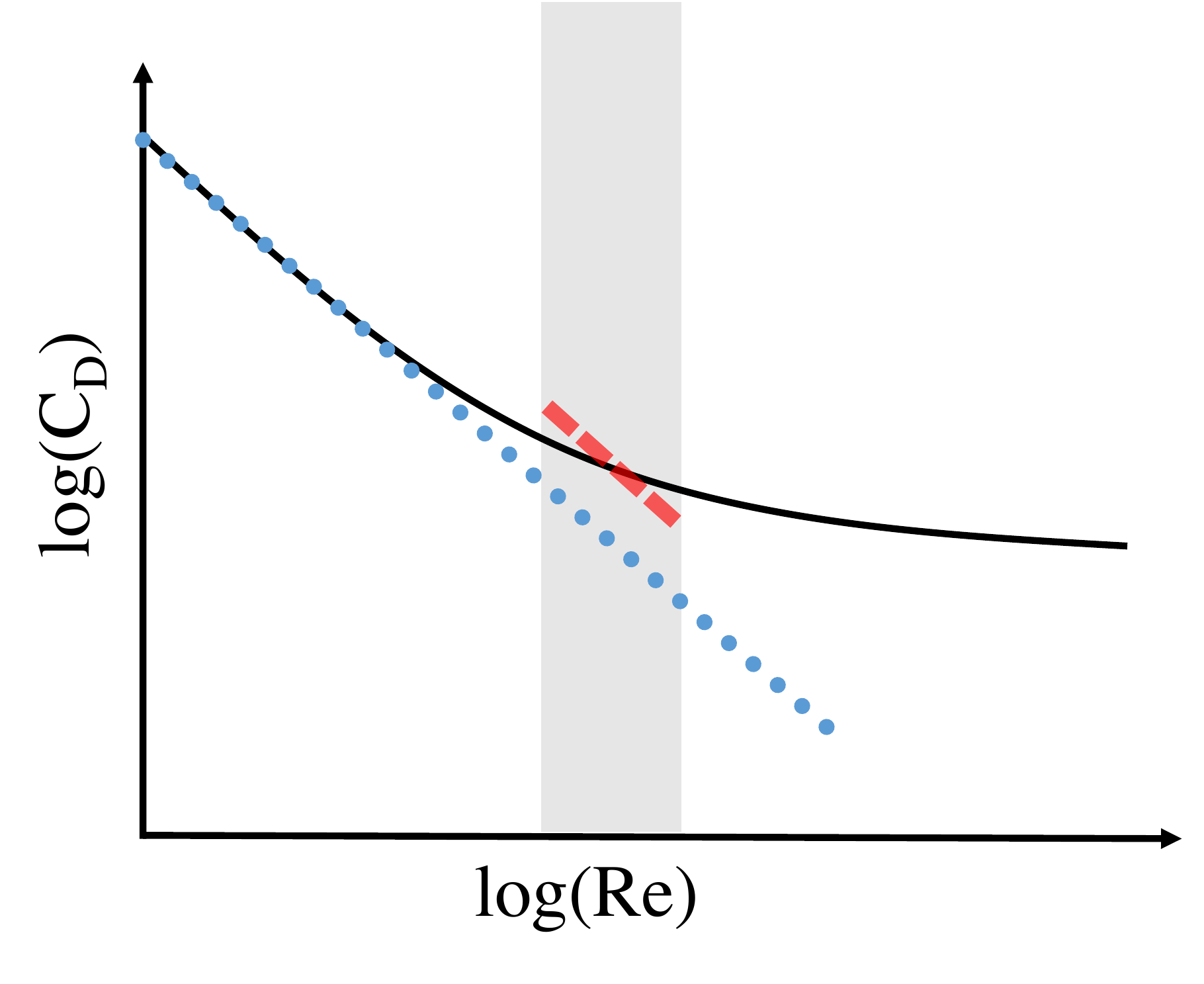}
            \caption{\label{Fig2} The black line indicates the coefficient of drag as a function of the Reynolds number for a particular shape.  The blue dotted line is the Stokes drag approximation for low $\Re$ and the red dashed line is the intermediate $\Re$ approximation.}
        \end{figure}

        The drag coefficient for a cylindrical disk will be a function of $\Re$ similar to that of a sphere.  The solid, black line in figure \ref{Fig2} indicates the coefficient of drag over a range of small to large $\Re$ for a general particle \citep{clift_bubbles_2013}.  The dotted blue line indicates the Stokes drag model, taking the form: $C_D = C_{1}/\Re$ where $C_1$ is a constant that will depend on the geometry of the particle ($C_{1}=24$ for a sphere and $C_{1}=13.6$ for a circular disk parallel to the flow) \citep{happel_brenner_1983}.   The Stokes model does not well represent the true value of $C_D$ in the shaded region, which is the intermediate $\Re$ regime that is typical of a swash flow.  Similar to \citet{dibenedetto_enhanced_2022}, we approximate $C_D$ in the intermediate $\Re$ regime as inversely proportional to $\Re$ with a new, unknown proportionality constant, $C_{2}$, noted by the dashed red line in figure \ref{Fig2}.  From equation \eqref{eq-particle_2}, we substitute $C_D$ for the intermediate $\Re$ regime to get   
        \begin{equation}
        \label{eq-accel-dimensional}
            \frac{dv'_p}{dt'}  = \beta \frac{Du'}{Dt'}  - \frac{(v'_p-u')}{\tau_p}  + \left[\; a'_F \;\right],
        \end{equation}
        where 
        \begin{equation}
        \label{eq-accel-parameters}
            \beta = \frac{\phi + C_m \phi}{\gamma + C_m \phi}; \quad
            \tau_p = \frac{\pi D_p^2 (\gamma +C_m \phi)}{2 C_{2} \phi \nu},
        \end{equation}
        are parameters that represent the strength of the fluid forcing and added mass of the fluid acting on the particle, and the relaxation time for the particle in the flow, respectively. Note, $\beta = 1$ when the water is deep enough to allow the particle to float and the unknown empirical coefficient in the drag formulation is absorbed into the particle timescale.

        In a swash flow event, the relevant scales of motion are the initial shoreline velocity ($U_s$) and gravity ($g$). Using these scales to make equation \eqref{eq-accel-dimensional} dimensionless, we define the dimensionless variables $u = u'/U_s$, $v_p = v_p'/U_s$, $t = t'/(U_s/g)$, and $a_F = a_F'/g$.  This gives
        \begin{equation}
            \frac{dv_p}{dt}  = \beta \frac{Du}{Dt}  + \frac{(u - v_p)}{\St}  + \left[\; a_F \; \right], \label{eq-accel-dimless}
        \end{equation}
        where the particle Stokes number is given by
        \begin{equation}
            \St = \frac{g \tau_p}{U_s}.
        \end{equation}   
        The fluid forcing and force from the added mass are captured in the first term and the drag of the fluid on the particle is captured in the second term of equation \eqref{eq-accel-dimless}. 
        
        In addition to the equation for particle motion, we use a classical model for swash flow \citep{shen_climb_1963, peregrine_swash_2001}. For a given beach slope $s$, the shoreline position, shoreline velocity, flow velocity field, and water depth throughout the swash zone are given by,
        \begin{subequations}
        \label{eq-nondim-swash}
        \begin{align}
            x_s &= t - \frac{1}{2}st^2, \label{eq-shoreline} \\
            u_s &= 1 - st,  \label{eq-shoreline-velocity} \\
            u &= \frac{1}{3} \left( 1 - 2st + 2\frac{x}{t} \right), \label{eq-fluid-velocity} \\
            h &= \frac{1}{9} \left( 1 - \frac{1}{2}st - \frac{x}{t} \right) ^2, \label{eq-water-depth}
        \end{align}
        \end{subequations}
        where $x_s = x_s'/(U_s^2/g)$ is the dimensionless shoreline position, $u_s = u_s'/U_s$ is the dimensionless shoreline velocity, $u = u'/U_s$ is the dimensionless fluid velocity, and $h = h'/(U_s^2/g)$ is the dimensionless water depth.  This bore-driven swash model has been shown to accurately describe the flow evolution in a single swash event in laboratory experiments \citep{pujara_swash_2015}.

    \subsection{Extra considerations for beaching particles} \label{sec-beaching-considerations}
        We supplement equation \eqref{eq-accel-dimless} to capture aspects of particle motion in the swash zone that are critical, but not included in the dynamics.  Specifically, we account for the dynamics of the swash tip as this is not captured in the swash model. We also consider the frictional force on the particle with the beach when the water depth goes to zero.
        
\begin{figure}
            \centering
            \includegraphics[width=0.6\textwidth]{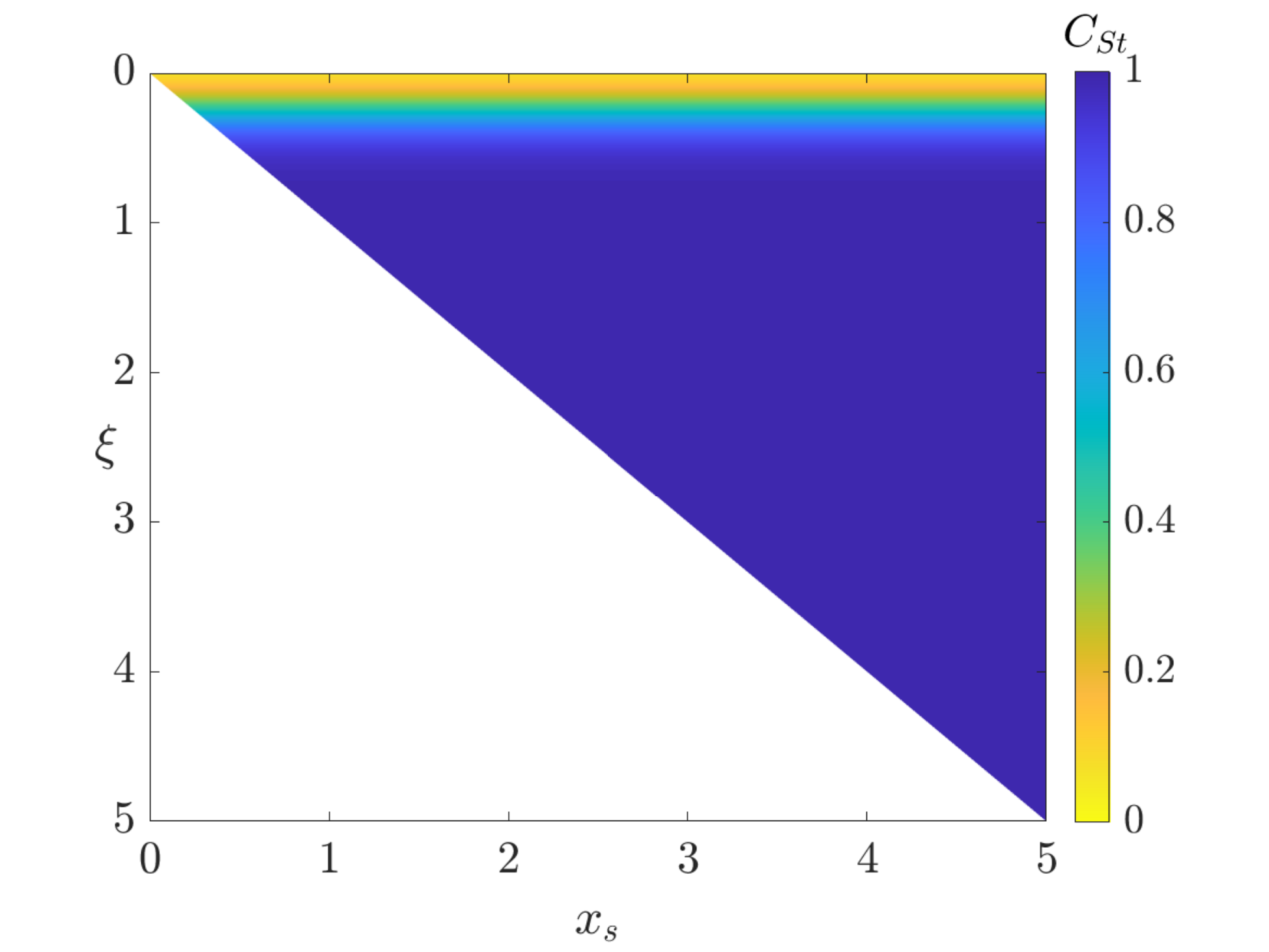}
            \caption{\label{Fig3} The Stokes number coefficient ($C_{\St}$) during the wave run-up, shown as a function of the particle position relative to the shoreline ($\xi$) and the shoreline position ($x_s$).  $C_{\St}$ decreases for a particle near the shoreline, while a particle far from the shoreline is unaffected.  The Stokes coefficient noticeably deviates from unity in the region of about 1/10 of the maximum run-up.}
        \end{figure}

        To account for particle motion in the swash tip, we first define the particle location ($x_p$) relative to the shoreline location ($x_s$) as
        \begin{equation}
            \xi = x_s - x_p.
        \end{equation}
        When the particle is behind the shoreline ($\xi > 0$), we determine the local fluid velocity and water depth from equations \eqref{eq-nondim-swash}.  In the model as it stands, if the particle is at the shoreline ($\xi = 0$), the water depth is zero and the fluid velocity becomes the shoreline velocity from equation (\ref{eq-shoreline-velocity}).  If the particle moves past the shoreline ($\xi < 0$), there is no fluid and thus the depth and fluid velocity at the particle position are both zero. The swash model does not account for the effects of friction and flow convergence at the swash tip, which has been shown to reduce shoreline velocity \citep{pujara_integral_2016} and collect particles during the wave run-up process \citep{baldock_flow_2014}. To account for this, we make two modifications. The first is that we expect the depth of the water in the swash tip to be greater than the particle (friction creates a `blunt nose' at the tip).  In this case, using equation \eqref{eq-gamma-phi} with the definition of $\beta$ from equation \eqref{eq-accel-parameters}, we make the simplification that $\beta=1$ throughout the wave run-up. Second, we implement a spatial dependency on the Stokes number during the run-up, which accounts for the fact that the particle becomes more tracer-like in the fast and turbulent swash tip,
        \begin{equation}
            \St_\text{eff} = C_{\St} \St = \frac{1}{1 + e^{-k(\xi-w)}} \St.
        \end{equation}
        Here, the coefficient $C_{\St}$ ranges from 0 to 1, acting to decrease the effective Stokes number in the vicinity near the tip during the wave run-up, which is essentially a boundary layer to the swash flow. The constant $k$ represents a sharpness factor that determines how smoothly the $C_\St$ varies with $\xi$, and $w$ determines the width of the zone where the $C_\St$ noticeably falls below unity. While the results presented below are not sensitive to the precise values, using $k = 10$ and $w = 0.25$ provides an effective region of about 1/10 of the maximum run-up where the Stokes number is modified (figure \ref{Fig3}). The swash tip size ($10\%$ of the maximum run-up) is an approximation, but consistent with previous data and models of the swash tip \citep{baldock_flow_2014, pujara_integral_2016}
        
        On the wave draw-down, we consider the particle to be prone to beaching if it is left behind by the fluid and the local water depth at the particle becomes zero.  In this instance, there is no buoyant force on the particle, but the particle's weight is balanced by the reaction force of the particle on the beach in the small-$\theta$ approximation introduced above.  In the $x$-direction, there is no longer any force from the fluid and the only force on the particle is the friction force acting opposite to its direction of motion.  We model this friction force using the standard formulation $F_F = \mu N = \mu m_p g$.  In the absence of water, the deceleration of the particle due to friction is then given by 
        \begin{equation}
            a_F = - \mu \sgn{}(v_p).
        \end{equation}

\subsection{Model results} \label{sec-model-results}

    The model equation \eqref{eq-accel-dimless} is solved in MATLAB using a standard solver (\textit{ode45}). The particle begins moving at $(x, t) = (0,t_{p0})$ with an initial velocity $v(x=0,t=t_{p0}) = V_p$. Initializing the model in this way also serves to avoid the singularity in the flow velocity, depth, and acceleration which are undefined at $(x=0,t=0)$. 

\begin{figure}
        \centering
        \includegraphics[width=0.85\textwidth]{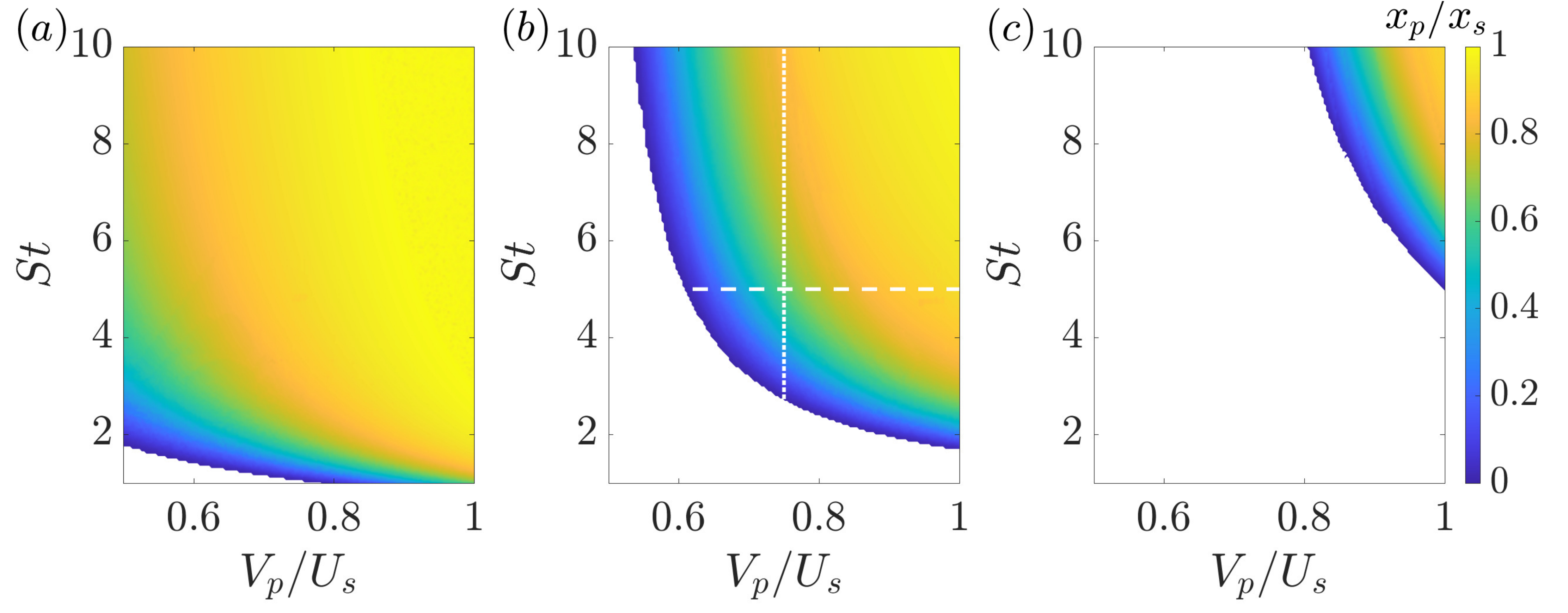}
        \caption{Fraction of final particle location over maximum shoreline run-up, for beaching particles as a function of particle Stokes number ($\St$) and dimensionless initial velocity ($V_{p}/U_s$) at three dimensionless initial particle times: (a) $t_{p0}/(2U_s/(gs)) = 0.001$, (b) $t_{p0}/(2U_s/(gs)) = 0.01$, (c) $t_{p0}/(2U_s/(gs)) = 0.1$, where $2U_s/(gs)$ is the duration of the swash event.  The dotted and dashed white lines indicate transects considered in figure \ref{Fig5}.}
        \label{Fig4}
    \end{figure}

    We show the impact of particle inertia and initial conditions on beaching dynamics in figure \ref{Fig4}.  The figure shows the final location of the particle at the given conditions.  After a particle is returned to the water, the model is no longer accurate and the final location is not plotted as the particle is not deposited on the beach.  We consider Stokes numbers in the range $\St \in [1,10]$, which is a realistic range considering that it covers $\tau_p \in [0.2, 2]$~s for a swash event with $U_s = 2$~m/s.  For the particle's initial velocity, we consider the range $V_p/U_s \in [0.5, 1]$. While the lower limit is an arbitrary choice, the upper limit corresponds physically to a particle that is caught in the bore front.  Apart from the initial velocity, the initial time at which the particle enters the swash zone is also important. The initial dimensionless time $t_{p0}/(2U_s/(gs)) = [0.001, 0.01, 0.1]$ can also be thought of as the distance that the particle starts behind the swash tip, where the smaller dimensionless time relates to the particle being closer to the bore front. Though we treat them separately in figure \ref{Fig4}, we note that the initial particle velocity and time are likely to be correlated such that a particle that enters the swash later is likely to do so at a lower velocity.  Additionally, we expect the particle $\St$ and the initial conditions to be related as the particle inertia likely dictates motion in the surf zone and thus the resulting initial conditions.

\begin{figure}
        \centering
        \includegraphics[width=0.8\textwidth]{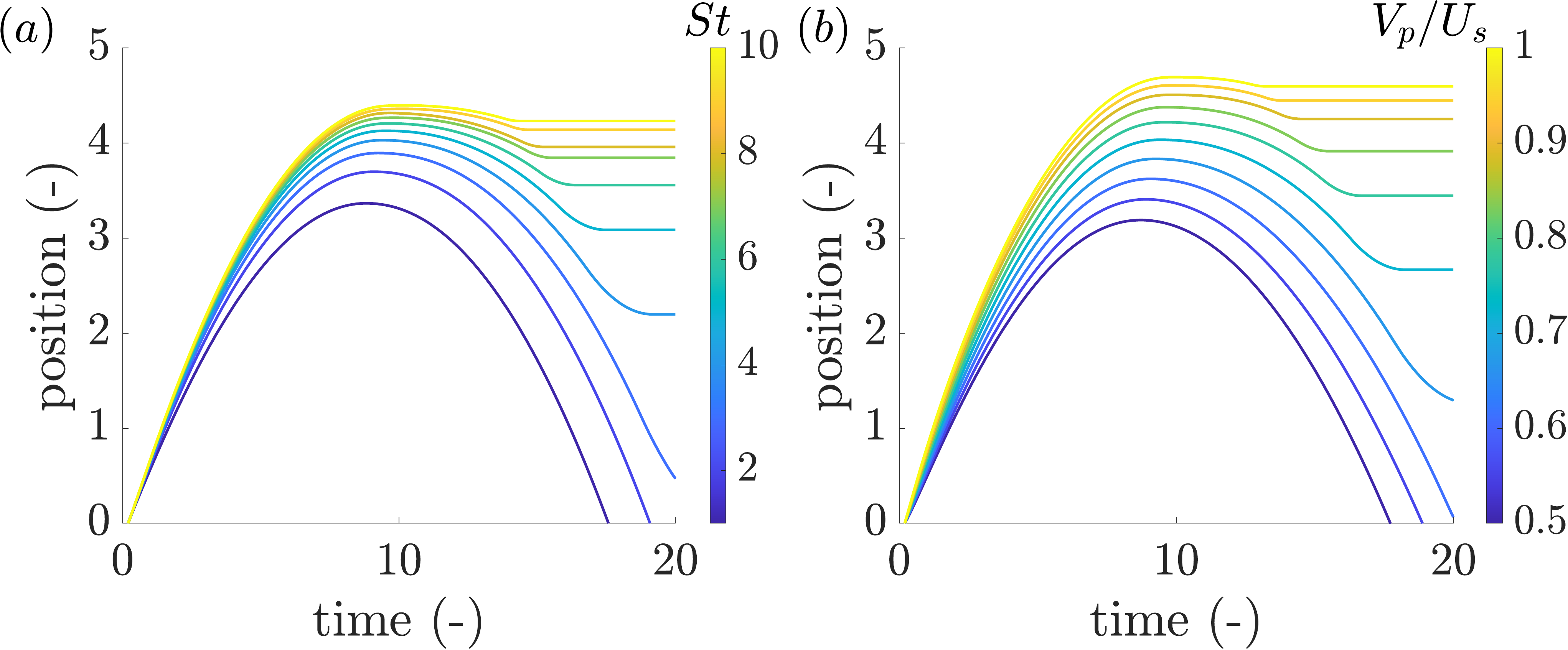}
        \caption{\label{Fig5} Model particle trajectories at transects from figure \ref{Fig4} with one varied parameter. a) Transect at the vertical dotted line in figure \ref{Fig4}b showing varied $\St$.  b) Transect at the horizontal dashed line in figure \ref{Fig4}b showing varied dimensionless velocity.}
    \end{figure}

    As expected, the likelihood of particle beaching increases at higher $\St$. This is because at higher $\St$, the particle does not change direction as quickly as the water. The water drains before the particle can respond, leaving the particle beached.  We also observe that particles are more likely to beach when they enter the swash zone at a higher initial velocity.  The larger initial velocity causes the particle to stay near the swash tip, reaching closer to the maximum wave run-up, and thus facilitating the particle to get caught on the beach during the downrush. When a particle enters the swash close behind the swash tip, we see that it is very likely to be beached (figure \ref{Fig4}a).  In this case, the beaching location is also very sensitive to $\St$.  When a particle enters the swash far behind the swash tip, it is far less likely to be beached (figure \ref{Fig4}c).  The particle will only beach in this case if it is very inertial and has a large initial velocity. Under the intermediate initial time (figure \ref{Fig4}b) beaching depends on both $\St$ and the initial particle velocity, which we explore further in figure \ref{Fig5}.

      Figure \ref{Fig5}a shows the particle trajectories from figure \ref{Fig4}b with $V_p/U_s = 0.75$ and varied $\St$.  We see that across this range of $\St$, the particle trajectory, and final particle location change significantly.  Similarly, figure \ref{Fig5}b shows the particle trajectories from figure \ref{Fig4}b with $\St = 5$ with varied initial particle velocities.  We again see that the particle trajectory and resultant beaching location change significantly.  Overall, whether a particle is beached and the location where it is beached is quite sensitive to the particle's swash Stokes number and its initial conditions as it enters the swash.

\section{Experiments}\label{sec-experiments}
\subsection{Wave flume experiments} \label{sec-experiment-setup}
We performed particle beaching experiments in the Water Science and Engineering Laboratory wave flume at the University of Wisconsin - Madison.  The wave flume is 39~m long with a 0.91~m width and 1.12~m height.  We use a piston-type wave maker to generate a solitary wave \citep{goring_tsunamis_1978} that acts incident to a 1/10 sloped impermeable, plane beach shown in figure \ref{Fig1}.  The generated wave has an amplitude of 0.06~m at the still water depth of 0.3~m ($a/h = 0.2$).  This wave incident on the specified beach produces a surging type breaking wave \citep{grilli_breaking_1997}.
    
For the particles, we place five plastic disks just offshore of the beach, near the still water line.  The particles are 3D printed from polypropylene. The particles have measured diameters of 2.261 $\pm$ 0.002~cm, heights of 0.268 $\pm$ 0.003~cm, and masses of 0.819 $\pm$ 0.002~g.  We calculate each particle density as the particle mass divided by the cylindrical volume from the respective height and diameter.  The average particle density is 0.760 $\pm$ 0.008~g/cm$^3$.  This is a bit less than the density of the raw polypropylene filament which is generally reported as 0.89~g/cm$^3$.
    
We use an obliquely mounted overhead digital camera with a frame rate of 30 fps to capture the swash event and particle motion. Throughout the camera field of view are calibration marks which are used to convert the particle location in pixels to the distance up the beach from $x = 0$. We track the particle trajectories and shoreline motion using custom code written in MATLAB.  For particle tracking, we first subtract the image background and binarize the image data based on a threshold that selects the particle.  We manually identify the starting location of each particle and for frames after the initial two frames, we find the particle velocity from the particle displacement over the previous two frames to predict the subsequent location of the particle.  We use the nearest neighbor algorithm from the predicted location to find the true location of the particle.

From the camera images, we also track the shoreline position which is clearly observed by the sharp wet/dry boundary of the wavefront moving up the beach.  We manually tracked the shoreline motion during the uprush by selecting the maximum run-up location in each frame.  We converted the location on the image from pixels to cm in the same way as for the particle tracking. 

At the very start of the swash cycle, the shoreline motion is obstructed from view by the steepness of the breaking wave as it collapses while passing through $x=0$. We trim the data to only consider the shoreline motion after the wave bore collapsed, and where the position of the shoreline could be accurately tracked. However, trimming the shoreline position data means we do not know the time at which the shoreline first moved nor its initial velocity, both of which are needed for the swash model. This uncertainty is also related to the physics of bore collapse. The swash solution in equations \eqref{eq-nondim-swash} is based upon the non-linear shallow water equations, in which the shoreline moves impulsively upon the wave bore arrival at the initial shoreline position.  In reality, the wave bore collapse process occurs over some finite time and distance \citep{pujara_swash_2015, yeh_bore_1988}. Since the bore collapse is not instantaneous, there is a brief time of shoreline acceleration before a maximum initial shoreline velocity is reached.

\begin{figure}
        \centering
        \includegraphics[width=0.5\textwidth]{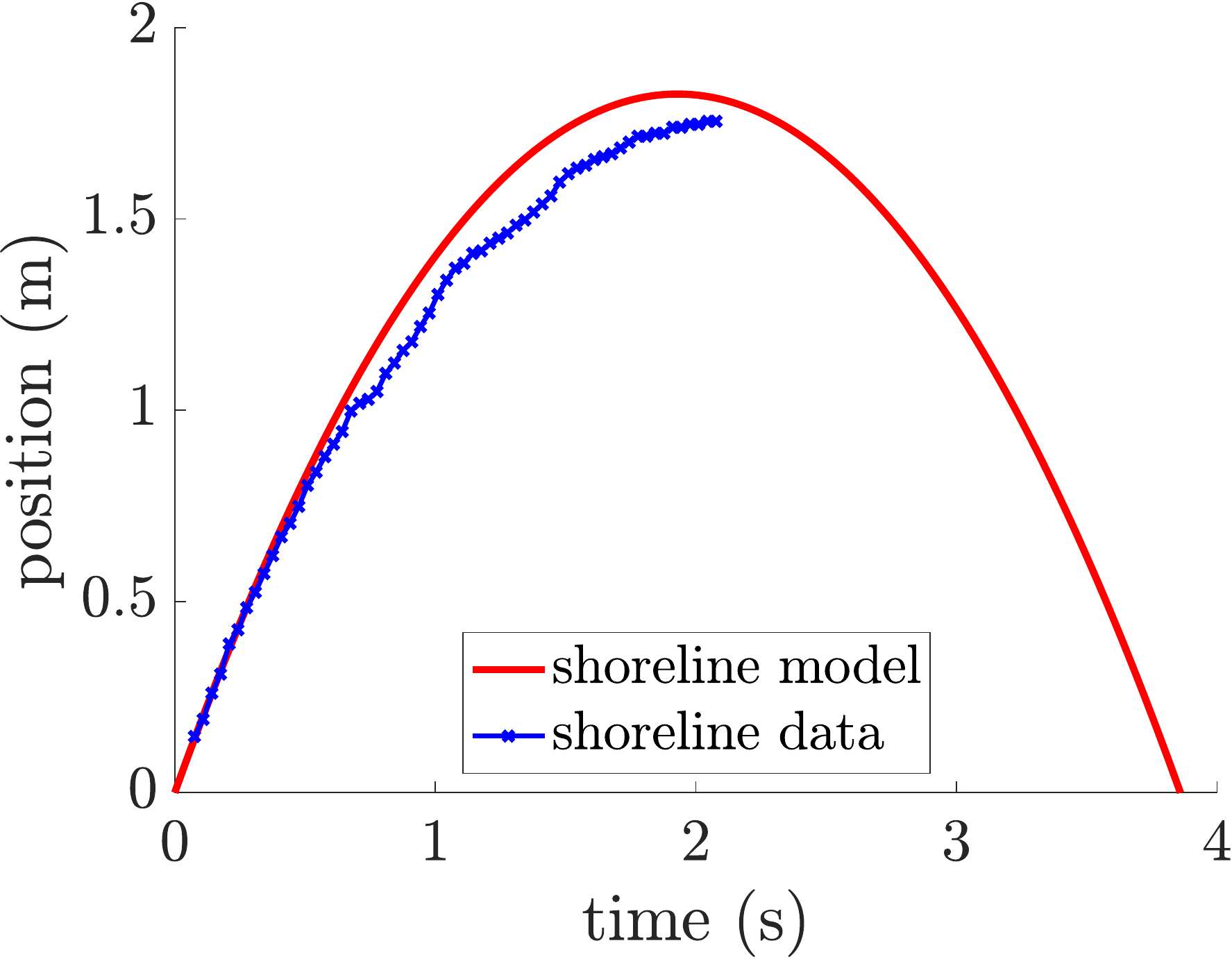}
        \caption{\label{Fig6} Run-up shoreline data and model fit to equation \eqref{eq-timeshift} with an initial shoreline velocity of $U_s = 1.89$~m/s and a time shift of $t_0 = 0.11$~s.}
\end{figure}

From the shoreline position data, we find the shoreline velocity using a Gaussian convolution \citep{ouellette_quantitative_2006, mordant_experimental_2004} with a filter width of 2 frames and  filter support of 7 frames.  In our data, we confirm that the maximum observed shoreline velocity occurs at the same time that the wave bore visually appeared completely collapsed. Since we expect the shoreline position after complete bore collapse to follow equation (\ref{eq-shoreline}) (at least in the absence of friction), and since $t=0$ is defined as the time at which the wave passes through $x=0$, we implement a time shift in the shoreline data to account for the finite time of the bore collapse process,
\begin{equation} \label{eq-timeshift}
    x_{s,\text{data}} = U_s(t-t_0) - \frac{1}{2}gs(t-t_0)^2.
\end{equation}
By fitting a portion of the data from the beginning of the wave run-up after bore collapse to equation (\ref{eq-timeshift}) using the maximum velocity after the bore collapse as an initial guess for the initial shoreline velocity, we find the initial shoreline velocity to be $U_s = 1.89$~m/s.  The shoreline data and model are plotted together in figure \ref{Fig6}, where we see that the shoreline model fits the shoreline data well at the start of the uprush.  The model predicts a slightly higher maximum run-up than we see in the experimental data, as expected since the shoreline model is frictionless. We do not have shoreline location during the draw-down from the experiment because tracking the shoreline location optically becomes near impossible during the draw-down process when there is no longer an obvious demarcation between dry and wet. 

The particle data are defined on the same time vector as the shifted shoreline data. To find the exact time and velocity with which each particle crossed the initial shoreline position, $t_{p0}$ and $V_p$, we fit a fifth order polynomial to each particle trajectory and analytically evaluate the zero crossing time and concurrent velocity.
        
\subsection{Experimental particle trajectories and comparison with model} \label{sec-exp-model-comparison}
\begin{figure}
        \includegraphics[width=1.0\textwidth]{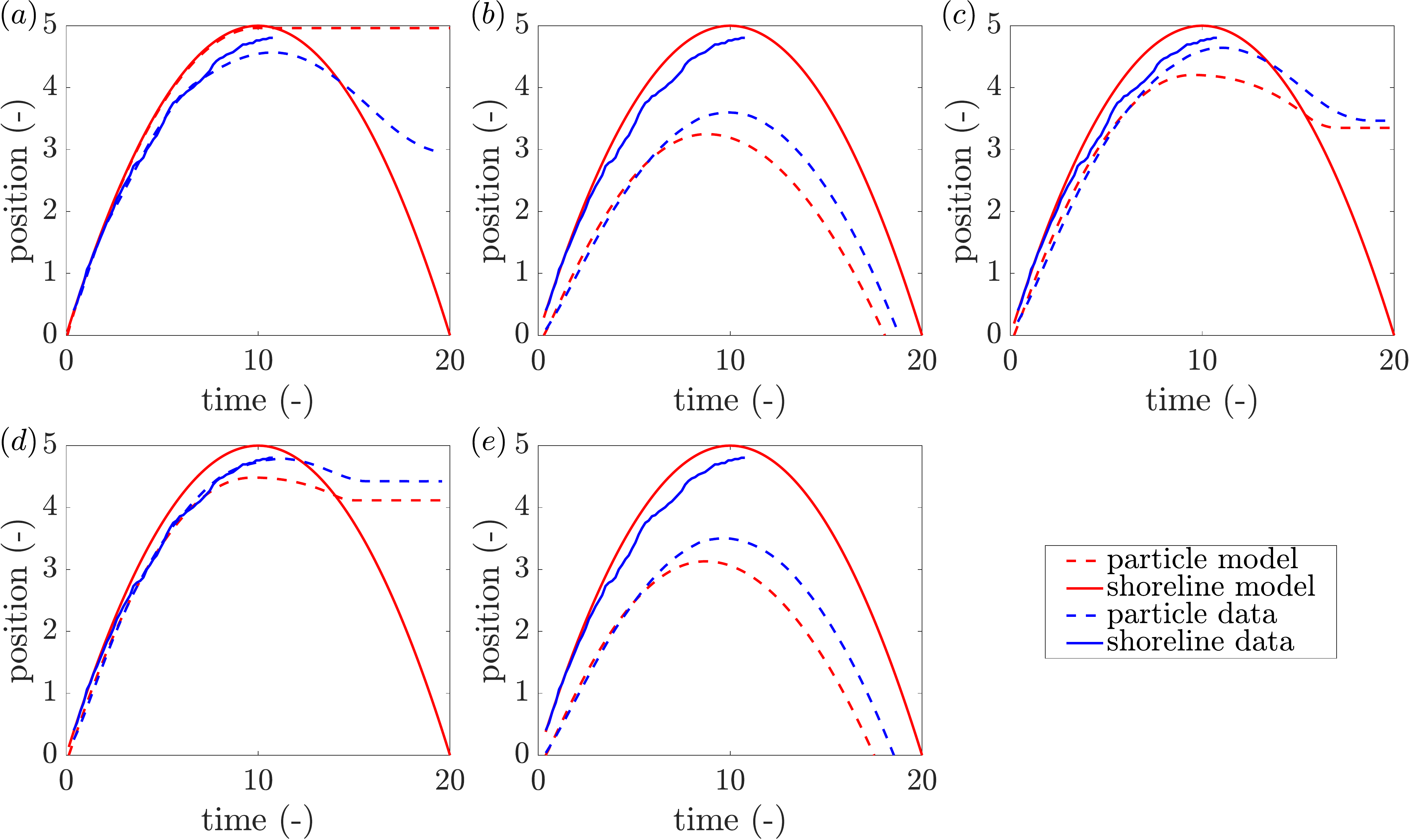}
        \caption{Model and experimental shoreline run-up plotted with model and experimental particle trajectory for five experimental particles with initial conditions listed in table \ref{table-initial-conditions}.  Figure panels a-e correspond to particles A-E respectively.}
        \label{Fig7}
    \end{figure}

\begin{figure}
        \centering
        \includegraphics[width=0.5\textwidth]{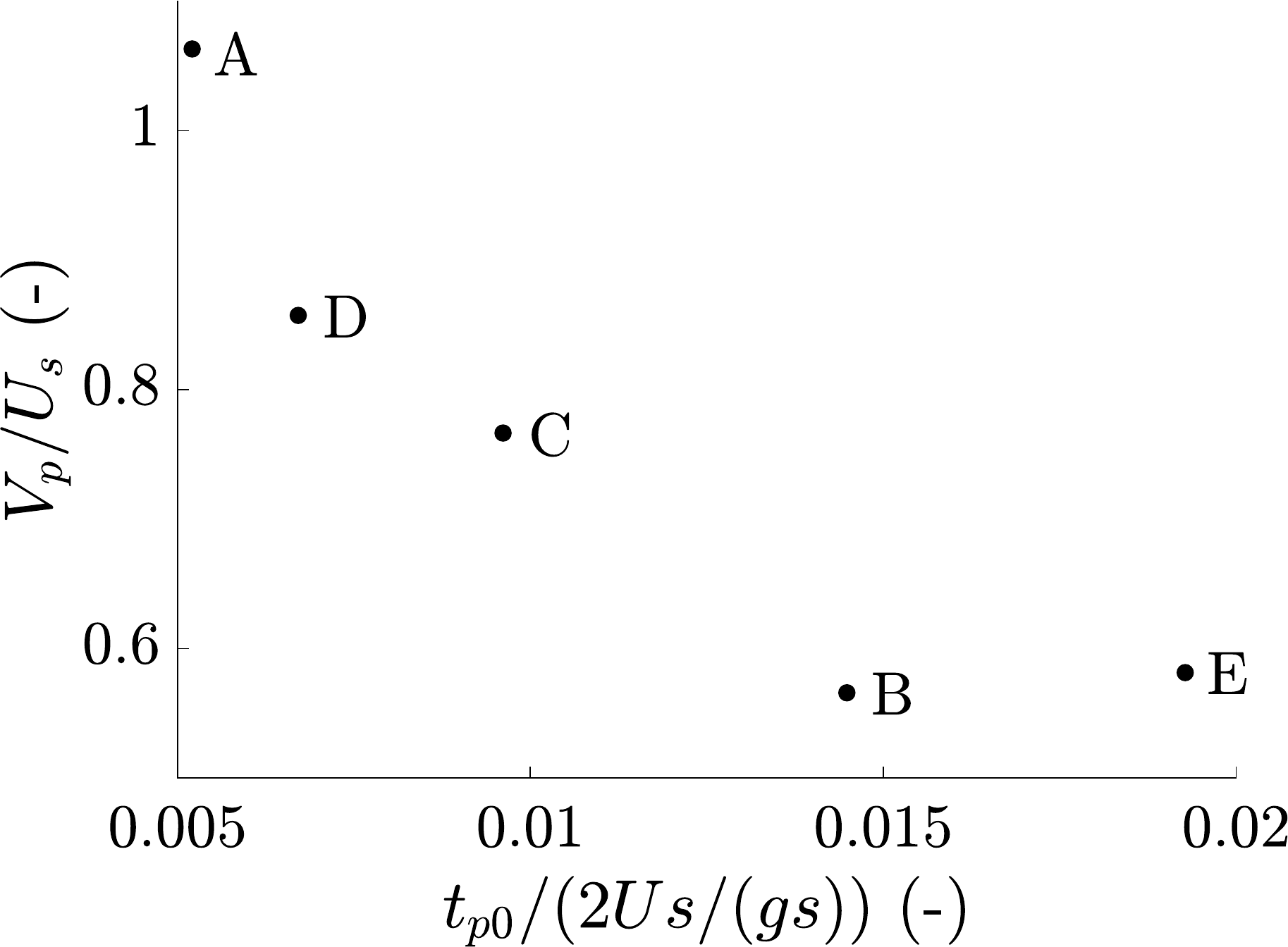}
        \caption{Initial particle velocity vs. initial particle time.  Particles that enter the swash zone closer to the wave will enter with a larger velocity, and those that are further behind the wavefront will enter the swash zone with a lower velocity.}
        \label{Fig8}
    \end{figure}

    In figure \ref{Fig7} we show the experimental particle and shoreline trajectories with the model particle and shoreline trajectories.  Table \ref{table-initial-conditions} lists the initial conditions for each particle and the particle fate.  The correlation between the initial particle velocity and initial particle time are shown in figure \ref{Fig8}.  Generally, a particle that enters the swash zone closer to the wavefront will enter with a greater velocity, compared to a particle that enters further behind the wave bore, which will enter with a lower velocity.  Besides the particle initial conditions, the rest of the model parameters are: $\tau_p = 1$~s (corresponding to $\St = 5.18$), $\gamma = 0.76$~g/cm$^3$, $C_m = 0.75$, $\mu = 0.15$, and $H_p = 0.268$~cm.  We choose $\tau_p$ as a reasonable order of magnitude guess of the particle timescale that facilitates agreement between the model and data, the values of $\gamma$ and $H_p$ are as measured from the experimental particles, and $C_m$ is taken as the average value between the added mass coefficient of a sphere ($C_m = 1/2$) and a circular cylinder ($C_m = 1$) \citep{newman_marine_book}.  The dynamic friction coefficient for non-metals usually ranges from 0.3 to 0.4 and is lower when moist or well lubricated \citep{rabinowicz_friction_1965}.  From the wetness of the beach surface after a swash event, we justify the estimated friction coefficient of $\mu = 0.15$.  We further explore the sensitivity of the model to these predicted values in $\S$ \ref{sec-discussion}.

    \begin{table}
        \centering
        \begin{tabular}{ |c|c|c|c|  }
             \hline
             Particle & $V_p/U_s$ & $t_{p0}/(2U_s/(gs))$ & Fate \\
             \hline
             A & 1.03 & 0.0028 & beach\\
             B & 0.57 & 0.015 & water\\
             C & 0.77 & 0.0096 & beach\\
             D & 0.86 & 0.0067 & beach\\
             E & 0.58 & 0.019 & water\\
             \hline
        \end{tabular}
        \caption{Experimental particle initial conditions and fate from figure \ref{Fig7}.}
        \label{table-initial-conditions}
    \end{table}

    The comparison between the data and model trajectories in figure \ref{Fig7} shows that the model captures the qualitative behavior observed in the experiments. In particular, the model trajectories correctly predict whether a particle is deposited on the beach (particles A, C, \& D) or returned to the water (particles B \& E) without any tuning of parameters. Of the beached particles, the location of beaching in the model is also close to the observed value in two of the three beaching particle cases (particles C \& D). The experimental particle trajectories match the model very well at the start of the swash event, with the deviation between the data and the model growing until the maximum run-up.  This shows the known limitations in the model setup: the simplified flow model (equations \eqref{eq-nondim-swash}) ignores the effect of friction in the flow, which accumulates over time as the flow reaches its maximum run-up. This is especially important near the swash tip, which is the reason the trajectory for particle A shows qualitative differences with the model.
        
    Apart from comparing the particle trajectories between the model and experiment, another useful comparison is considering the particle trajectory relative to the shoreline trajectory in the experiment compared with the same thing in the model.  This captures whether the model contains enough physics to predict particle beaching dynamics. In this regard, we see that the trends of particle behavior with the shoreline compare very well for all three beaching particles in the experiment and the model. Not only do the model trajectories capture the particle deceleration during uprush that is the result of the complex interplay between the unsteady flow and particle drag, but they also capture how the particle decelerates to zero velocity as they get beached. 
        
    Overall, the particles that were deposited on the beach surfed on the wavefront during the uprush, while the non-beaching particles were behind the wave from the very beginning. Given that all particles were the same in this experiment, the fact that beaching was very sensitive to initial conditions is consistent with the model results ($\S$ \ref{sec-model-results}).
        
\begin{figure}
         \includegraphics[width=1.0\textwidth]{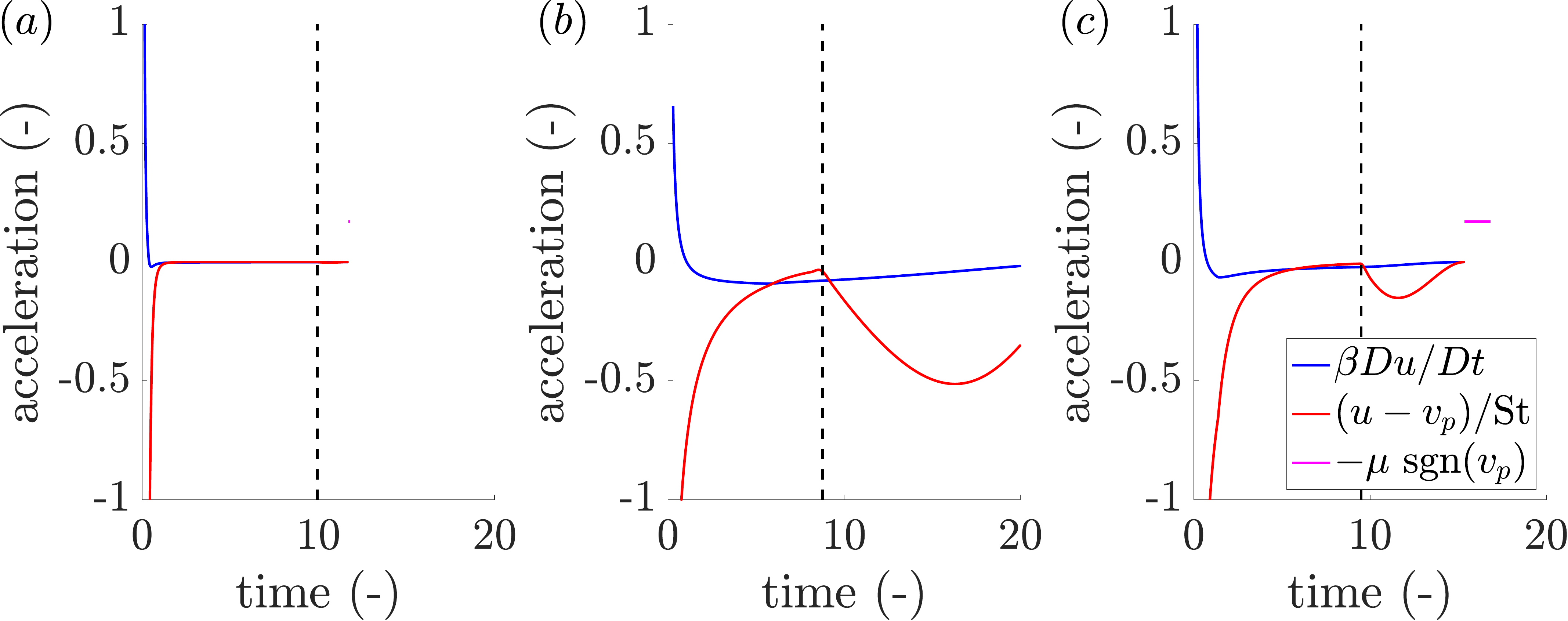}
        \caption{Acceleration components actively contributing to the particle force balance as a function of time for particles A-C from figure \ref{Fig7}(a-c) respectively.  The dashed black line indicates the point in time where the fluid at the particle location changes direction.  We note that there is a small pink region in (a), none in (b), and a larger pink region in (c).}
        \label{Fig9}
    \end{figure}

    To further understand the dynamics of each particle, we consider the acceleration budget of particles A, B, \& C, in figure \ref{Fig9}.  Two of these are beaching particles (particle A, deposited high on the beach in the model, and particle C, deposited lower on the beach in the model), and one returns to the water (particle B).  At the beginning of each particle's trajectory, each particle experiences a positive (up the beach) force from the first term in equation \eqref{eq-accel-dimless} and a greater force acting in the opposite, negative (down the beach) direction from the middle term of equation \eqref{eq-accel-dimless}.  For the particles that eventually beach, these two forces both approach zero quickly and stay there, although faster for particle A (figure \ref{Fig9}a) which is beached close to the maximum run-up.  For the particle that returns to the water (particle B, figure \ref{Fig9}b), the first and second terms decrease in magnitude slower.  After the flow reverses direction at the particle location, noted by the vertical black dashed line, the second term (drag) increases in magnitude. The increased drag force is in the negative direction acting to wash the particle back into the water.  We see this slightly in particle C (figure \ref{Fig9}c), but the particle runs out of water before it travels sufficiently downslope to return to the main water body, resulting in particle beaching.

\section{Discussion}\label{sec-discussion}
The main free parameters in this model are the particle initial conditions and the particle Stokes number, $\St$.  The dimensionless initial particle time and velocity are characteristics of the particle motion from the surf zone into the swash zone; they are consequences of the particle forcing within the surf zone and are related as we showed in figure \ref{Fig8}.  The particle Stokes number $\St$ is a measure of the particle inertia and thus a metric of the particle itself that in turn corresponds to a representative drag model.  We considered the impact of varied $\St$ and initial conditions in figures \ref{Fig4} and \ref{Fig5}, and now we consider the impact on the model from these and the additional parameters $C_m$, $\mu$, $\gamma$, $H_p$, and $s$.  We start by considering a single model particle trajectory with $\St = 5$, $V_p/U_s = 0.75$, and  $t_{p0}/(2U_s/(gs)) = 0.01$.  This trajectory is noted by the intersection of the dotted and dashed lines in figure \ref{Fig4}b.  We then solve for the trajectory of this particle while changing individual parameters over a range of reasonable values.  We examine the spread of the final particle location due to the parameter deviations.  In table \ref{table-varied_parameters}, we list the parameters, the range of examined values, and the resulting range of the final particle beaching location.  Here we note that the swash model becomes inadequate after the shoreline surpasses the initial location on the downrush, so particles that fall beyond the initial shoreline position are characterized as non-beaching and noted with a final location of 0.  The parameters are listed in descending order of importance on model output.  We see that the parameters that have the most impact are the physical parameters already discussed above: the particle initial conditions ($t_{p0}$, $V_p$), and the particle inertia ($\St$). An additional significant parameter is the beach slope, which was not varied in these experiments but appears to affect the particle beaching location significantly.  The particle height also did not change in the experiments but shows to have a moderate effect on the particle beaching. The impacts from changing $\mu$, $\gamma$, and $C_m$ are an order of magnitude smaller than the other parameters, thus the estimated values are sufficient for approximating the particle trajectory.
    \begin{table}
        \centering
        \begin{tabular}{ |c|c|c|  }
             \hline
             Parameter & Estimated Range & Final Location Range \\
             \hline
            s & 1/20 - 1/5 & 1.5 - 6.2 \\
            $V_p/U_s$ & 0.5 - 1 & X - 4.6 \\
             $t_{p0}/(2U_s/(gs))$ & 0.001 - 0.1 & X - 4.6 \\
             St & 1 - 10 & X - 4.2 \\
             $H_p$ & (0.1 - 0.5) cm & 2.0 - 3.4 \\
             $\mu$ & 0.1 - 0.5 & 3.1 - 3.3 \\
             $\gamma$ & (0.7 - 0.9) g/cm$^3$ & 3.1 - 3.3 \\
            $C_m$ & 0.5 - 1 & 3.1 - 3.2 \\
             \hline
        \end{tabular}
        \caption{Model parameters with an estimated range of appropriate values and the resulting difference in final particle location when varied across the specified range.  The parameters are listed in descending order of the final particle location range.  X notes that the particles were returned to the water.}
        \label{table-varied_parameters}
    \end{table}

We can also determine in which direction each parameter impacts the particle's beaching location.  As already pointed out in figure \ref{Fig4}, decreases in the initial time increases the particle location up the beach while increases in the initial velocity and $\St$ both increase the final location of the particle.  For the other parameters, the final beaching location increases with increases to $H_p$, $\mu$, and $\gamma$, and decreases with increases to slope and $C_m$.  Although we are only considering the trajectory of one model particle here, this sensitivity analysis helps us to understand how these parameters can change the model result in both magnitude and direction.

We have simplified this model by making certain assumptions about the swash flow and interaction with particles.  First, in the given swash model since the bore front is not accurately captured, we implement $\beta = 1$ on the uprush.  This is reasonable due to the bore front, where in reality, the water depth will be greater than the particle height at the swash tip for most of the run-up. Additionally accounting for the swash tip, we implement the effective Stokes number scheme.  The parameters selected are reasonable and facilitate an agreement of the experimental results with the model. Concerning the implementation of friction with the beach, here we take a simple approach that friction starts once the beach is dry.  However, as soon as the water depth decreases such that the particle is in contact with the beach, the force of friction would begin acting on the particle.  This may require a transitional regime where friction with the beach begins before the beach is completely dry at the particle location, which would cause the change in forces to be more gradual.  Although we make these assumptions and suggest such improvements, these assumptions hold for our simple model, demonstrated by the agreement with experiments and the analysis in $\S$ \ref{sec-exp-model-comparison}.

\section{Conclusions}\label{sec-conclusions}

We have derived a solution for buoyant, inertial particle transport through bore-driven swash to understand the factors leading to marine debris beaching.  This is not an exhaustive model, but rather a minimal model that includes only the most essential dynamics, which can be used to predict particle beaching effectively.   We have validated this model with experiments and explored the sensitivity of its results to the parameterizations and assumptions made.  The particle initial conditions entering the swash zone and the particle inertia are the most important factors dictating the particle fate in regard to beaching; the particle velocity and location leading into the swash zone dictate whether the particle catches and surfs in front of the wave, while the particle inertia dictates whether the particle will become beached as a result of being unable to change direction and follow the fluid down the slope.  At present, this model considers only cross-shore transport and the idealized case of a single-bore driven swash on a plane beach.  Complexities will arise when accounting for oblique waves, multiple bore wave trains, debris of different shapes where orientation dynamics are important, debris of larger size where lubrication pressures may become important, and other environmental factors such as natural beach morphology and wind effects.  However, we do expect that the essence of the debris-beaching process we have identified will hold regardless of additional parameters and environmental factors.

\paragraph{Acknowledgements}
We acknowledge fruitful conversations with Gautier Verhille.

\paragraph{Funding Statement}
We gratefully acknowledge funding from the Freshwater Collaborative of Wisconsin (Grant No. T2-5/20–06).

\paragraph{Declaration of Interests}
The authors declare no conflict of interest.

\paragraph{Author Contributions}

B.D., J.B., and N.P. developed the model;  B.D. and N.P. designed the experiments; B.D. performed the experiments; B.D. wrote the manuscript; N.P. and J.B. provided revisions; N.P. supervised the project.

\bibliographystyle{apacite}
\bibliography{buoyantbeaching_arXiv.bib} 

\end{document}